# Actively tunable terahertz electromagnetically induced transparency analogue based on vanadium-oxide-assisted metamaterials

Zhaojian Zhang [1], Junbo Yang [2,*], Yunxin Han [2], Xin He [2], Jingjing Zhang [1], Jie Huang [1], Dingbo Chen [1], Siyu Xu [1] and Wanlin Xie [1]

[1] College of Liberal Arts and Sciences, National University of Defense Technology, Changsha 410073, China;

[2] Center of Material Science, National University of Defense Technology, Changsha 410073, China;

*Correspondence: yangjunbo@nudt.edu.cn

**Abstract:** Recently, phase-change materials (PCMs) have drawn more attention due to the dynamically tunable optical properties. Here, we investigate the active control of electromagnetically induced transparency (EIT) analogue based on terahertz (THz) metamaterials integrated with vanadium oxide ($VO_2$). Utilizing the insulator-to-metal transition of $VO_2$, the amplitude of EIT peak can be actively modulated with a significant modulation depth. Meanwhile the group delay within the transparent window can also be dynamically tuned, achieving the active control of slow light effect. Furthermore, we also introduce independently tunable transparent peaks as well as group delay based on a double-peak EIT with good tuning performance. Finally, based on broadband EIT, the active tuning of quality factor of the EIT peak is also realized. This work introduces active EIT control with more degree of freedom by employing $VO_2$, and can find potential applications in future wireless and ultrafast THz communication systems as multi-channel filters, switches, spacers, logic gates and modulators.

## 1. Introduction

Over the past decades, Metamaterials (MMs) have been focused continually due to the capability to manipulate electromagnetic (EM) waves in an unnatural way [1]. By designing artificial meta-resonators of MMs and arranging them appropriately, MMs can tailor lightwaves in subwavelength scale, consequently providing optical responses with desirable properties. In recent years, MMs have come to the terahertz (THz) regime [2]. Located between infrared and microwave band, THz radiations have enjoyed a rise of interest and are promising for security scanning [3], future wireless communications as well as the sixth-generation (6G) networks [3-5]. Now, MMs have been regarded as ideal platforms to achieve chipscale THz devices, such as THz sources [6,7], modulators [8,9], sensors [10,11] and absorbers [12,13].

Recently, electromagnetically induced transparency (EIT) analogue in THz MMs have attracted more attention [14-16]. EIT refers to a sharp transparent window within a broad absorption spectrum, which comes from the quantum interference between two distinct excitation pathways in the natural three-level atom system [17]. Due to the dispersion properties, EIT has potential applications in slow light, optical data storage, nonlinear process enhancement and signal processing [18]. Mimicking EIT in the classical system, MMs can reproduce such effect via the near-field coupling between bright and dark modes supported on meta-resonators [14]. Compared with the conventional EIT, which requires severe experimental conditions [18], MM-based EIT analogue is easier to be produced and more stable, therefore suitable for practical chipscale applications [15].

At the same time, for THz communication systems, THz devices with active tunability are required [19]. Therefore, various active optical materials have been utilized to turn passive MMs to be active, including liquid crystals [20,21], semiconductors [19,22,23], two-dimension (2D) materials [24-26] and phase-change materials (PCMs) [27]. PCMs, such as chalcogenide GeSbTe (GST) and vanadium dioxide ($VO_2$), possess significantly different optical properties in different phase states, i.e., amorphous and crystalline states [28]. Especially, $VO_2$ undergoes an insulator-to-metal transition when reaching its phase-change condition, which leads to a considerable increase in conductivity [29]. This phase transition can be triggered via thermal, electrical and optical ways, and it is also reported that such process has a sub-picosecond (sub-ps) response time under optical stimulation, therefore can meet the requirement of ultrafast THz interconnects [30,31]. Up to now, $VO_2$ has been integrated into THz MMs to realize various active functions, such as transmission control [30, 32-34], phase modulation [35-38], tunable absorbers [39,40] and switchable multifunctional devices [41,42].

In this paper, we numerically propose the active control of EIT analogue based on a vanadium-oxide-assisted THz metamaterial. Different from the previous work [43], in which $VO_2$ is used as the substrate, in this work, $VO_2$ islands are embedded into the meta-resonators. Via this strategy, the amplitude of EIT peak can be actively switched with a significant modulation depth utilizing the insulator-to-metal transition of $VO_2$. At the same time, the group velocity of light within the transparent window can also be dynamically tunable, achieving the active control of slow light effect. Furthermore, via coupling multiple meta-resonators, double and broadband EIT are investigated, and the transition of $VO_2$ will lead to the independent tunability of double EIT peak and broad-to-narrow band EIT evolution, respectively. The corresponding slow light dynamics are also investigated. This work achieves EIT modulations with multiple degree of freedom, and will find potential applications in the future THz wireless communication systems as multi-channel filters, switches, spacers, logic gates and modulators.

## 2. Structures, materials and methods

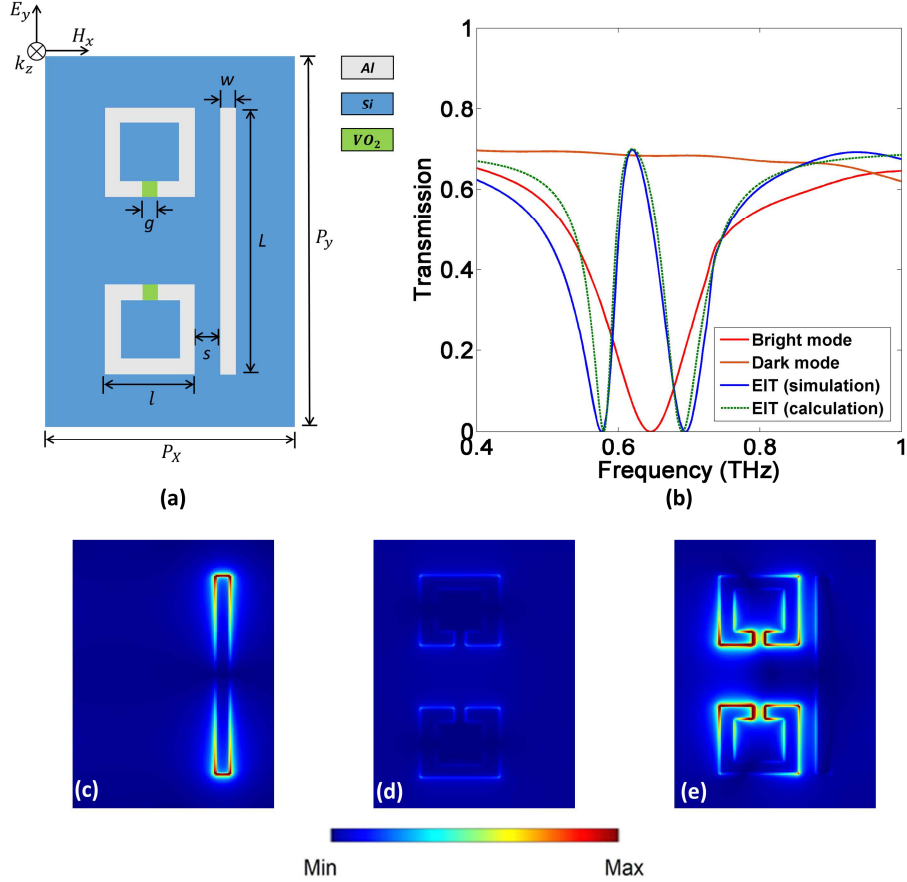

Fig. 1 (a) The top view of the proposed MM. The geometric parameters are as follows: *w*= 5 um, *L*= 85 um, *s*= 7 um, *g*= 5 um, *l*= 29 um, $P_x$= 80 um, $P_y$= 120 um. The height of CWR and SRRs (in *z* direction) is 200 nm. (b) The simulated transmission spectra of bright mode, dark mode as well as EIT and calculated transmission spectrum of EIT. (c) The |***E***| distribution of only CWR under y-direction polarized light at 0.65 THz. (d) The |***E***| distribution of only SRRs under y-direction polarized light at 0.65 THz. (e) The |***E***| distribution of CWR side-coupled with SRRs under y-direction polarized light at 0.65 THz.

The top view for the unit cell of proposed VO$_2$-embedded MM is illustrated in Fig. 1(a), and corresponding geometric parameters are given in the caption of Fig. 1. The unit cell includes double split-ring resonators (SRRs) side-coupled with one cut-wire resonator (CWR). Resonators are all made of aluminum (Al) and the gaps of SRRs are both embedded with VO$_2$. The substrate is chosen as silicon (Si) and is assumed to be semi-infinite in the *z* direction. The unit cell is periodic along *x* and *y* direction. Here, Al is described as a lossy metal with the conductivity $\sigma_{Al} = 3.72 \times 10^7$ S/m, and the refractive index of Si is set as $n_{Si} = 3.42$ with no loss [26,44]. The relative permittivity of VO$_2$ is expressed by Drude model [41]:

$$\varepsilon(\omega) = \varepsilon_\infty - \frac{\omega_p^2 \frac{\sigma}{\sigma_0}}{\omega^2 + i \cdot \omega \cdot \omega_d} \qquad (1)$$

where the permittivity at the infinite frequency $\varepsilon_\infty = 12$, the bulk frequency of plasma $\omega_p = 1.4 \times 10^{15}$ s$^{-1}$, the damping frequency $\omega_d = 5.75 \times 10^{13}$ s$^{-1}$ and $\sigma_0 = 3 \times 10^5$ S/m. Specially, the conductivity $\sigma$ of VO$_2$ can continuously increase by $10^4$ - $10^5$ orders of magnitude when VO$_2$ undergoes an insulator-to-metal transition. Such phase-change transition can be realized by thermal, electrical and optical stimuli. VO$_2$ has a gentle phase-change temperature at 340 K approximately, but thermal and electrical (which also attributes to the thermal effect) method only possess sub-second switching time [30]. However, sub-ps response time can be expected under the optical stimulus [31]. Here, we set $\sigma = 10$ S/m and $10^5$ S/m when VO$_2$ is in its insulating and fully metallic state, respectively [32,45].

The numerical simulation is investigated via finite-difference time-domain (FDTD) method. In the simulation, the periodical boundary conditions are adopted in the *x* and *y* directions and perfectly matched layers (PMLs) are employed in the *z* directions. Moderate mesh size is employed to ensure the simulation accuracy. A THz plane wave source with *y*-direction polarization and *z*-direction injection is set on the top of the MM, and a power monitor is put in the lossless substrate to detect the transmission.

### 3. Results and discussion

At first, EIT analogue is numerically investigated based on the proposed MM without VO$_2$. When only CWR exists in the unit cell, the y-direction polarized light can stimulate the dipolar resonance supported by CWR at 0.65 THz, such directly excited mode by the source is called bright mode. When there are only double SRRs, SRRs can support inductive–capacitive (LC) resonances at the same resonant frequency, but cannot be directly stimulated by the source with such polarization, therefore serving as dark modes. The corresponding transmission spectra as well as distributions of electric field magnitudes (|*E*|) at 0.65 THz are shown in Fig. 1(b-d). However, when double SRRs are placed aside CWR, LC resonances of SRRs can be excited by the dipolar resonance supported on CWR. Therefore, the near-field coupling between bright and dark modes will bring about the destructive interference, leading to a transparent peak in the transmission spectrum at 0.62 THz as shown in Fig. 1(b), the corresponding distribution of |*E*| is depicted in Fig. 1(e). Such phenomenon is known as EIT analogue, and can be quantitatively described by a coupling harmonic oscillator model. All the theoretical details will be discussed in section 4, and the result of quantitative calculation is also plotted in Fig. 1(b), showing a good fitting with simulated result.

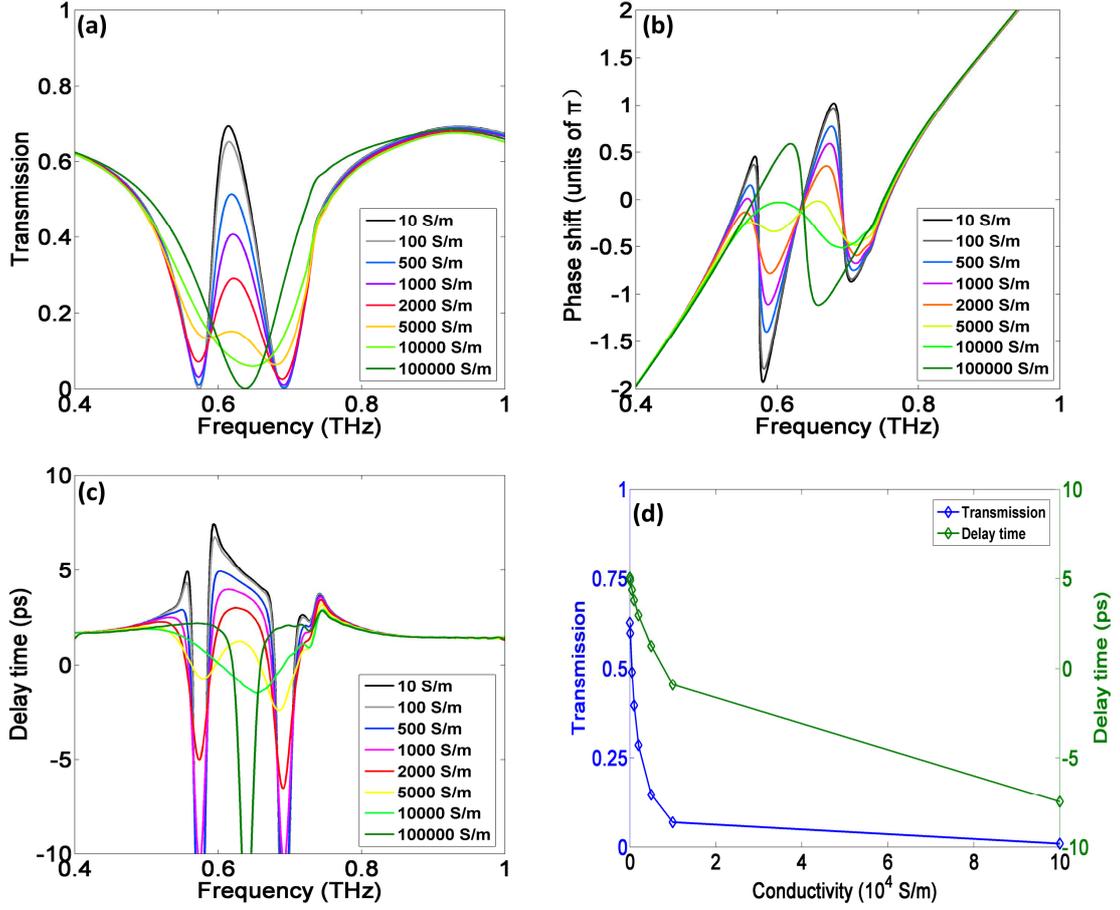

Fig.2 (a) The transmission spectra of EIT under different conductivities of $VO_2$. (b) The phase shift under different conductivities of VO2. (c) The delay time under different conductivities of $VO_2$. (d) The transmission as well as delay time at 0.63 THz under different conductivities.

Next, $VO_2$ islands are embedded in the gaps of the two SRRs as shown in Fig. 1(a), and transmission spectra under different conductivities of $VO_2$ are presented in Fig. 2(a). When $VO_2$ is in the insulating state with conductivity 10 S/m, the transparent peak is significant. However, as $VO_2$ transforms to the metallic state, the increasing conductivity will lead to a continuous decline of the peak amplitude. When $VO_2$ comes to its fully metallic states with conductivity $10^5$ S/m, EIT is totally suppressed. Meanwhile, the dispersion modulation will be introduced within the transparent window. Therefore, the phase shift of transmitted EM waves, defined as the phase difference between the power monitor and the source, is also dramatically modified as shown in Fig. 2(b), which will also cause the change of group velocity. The group delay, i.e., the slow light effect, can be assessed by the delay time [44]:

$$\tau_g = \frac{d\psi(\omega)}{d\omega} \qquad (2)$$

where $\tau_g$ is the delay time, and $\psi(\omega)$ stands for the phase shift. The delay time corresponding to different conductivities of $VO_2$ is shown in Fig. 2(c). Obviously, as the peak of EIT goes down, the delay time within the transparent window will drop simultaneously. To evaluate the modulation quantitatively, the transmission as well as delay time at 0.63 THz under different conductivities are plotted in Fig. 2(d). Utilizing the phase transition of $VO_2$, the transmission at 0.63 THz can be switched from 63% to 0.97%. The modulation depth (MD), defined as $(T_i - T_m)/T_i$ [46], where $T_i$ and $T_m$ are the transmission at 0.63 THz corresponding to insulating and fully metallic states of $VO_2$ respectively, is 98.46%. At the same time, the delay time can be manipulated from 5.06 ps to be negative. Therefore, active control of the transmitted THz radiation intensity as well as group velocity can be achieved by such structure with significant performance.

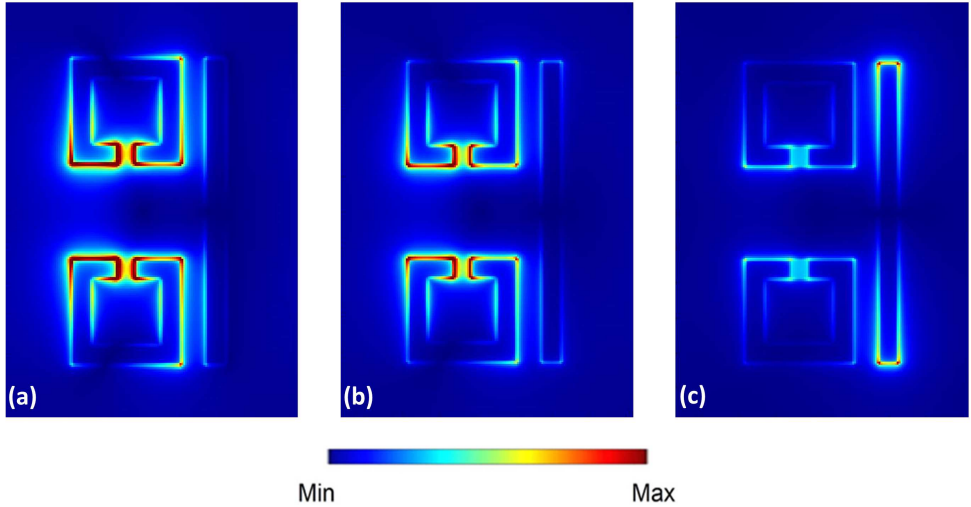

Fig. 3 (a) The |***E***| distribution at 0.63 THz when conductivity of $VO_2$ is 10 S/m. (b) The |***E***| distribution at 0.63 THz when conductivity of $VO_2$ is $10^3$ S/m. (c) The |***E***| distribution at 0.63 THz when conductivity of $VO_2$ is $10^5$ S/m.

To investigate the physical mechanism behind such modulation, the distributions of |***E***| at 0.63 THz with 10 S/m, $10^3$ S/m and $10^5$ S/m are shown in Fig. 3. Apparently, LC resonances supported on SRRs are constantly suppressed with the increasing conductivity. This is because the gap of SRR acts as the capacitor, and more loss will be introduced to the capacitor as $VO_2$ comes to be metallic, consequently weaken the capacitance as well as the LC resonances. As a result, the coupling from LC to dipolar resonances become weaker, leading to the vanishment of EIT effect.

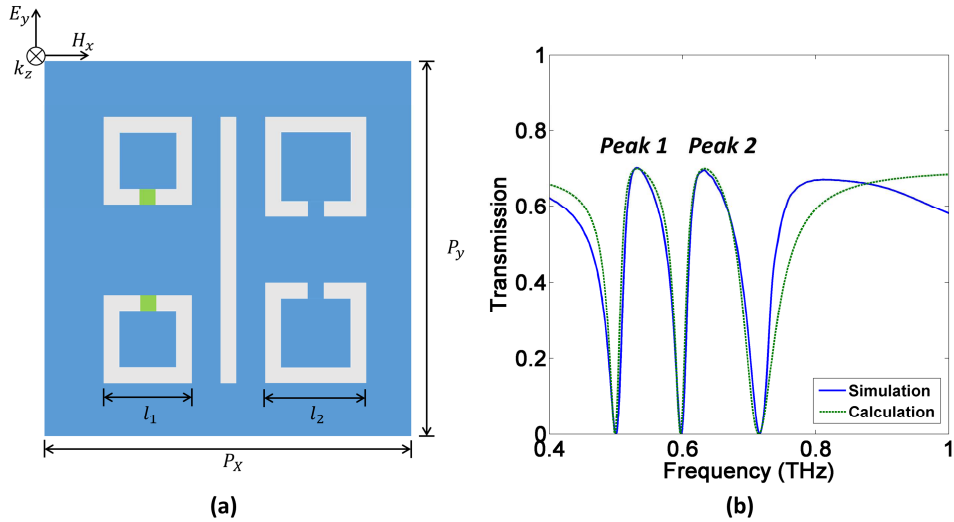

Fig. 4 (a) The structure to realize EIT with double peaks. The geometric parameters are unchanged except $l_1$= 28 um, $l_1$= 32 um, $P_x$= 116 um and $P_y$= 120 um. (b) The simulated and calculated transmission spectra of EIT with double peaks.

Furthermore, the EIT with double independently tunable peaks is also proposed based on the structure shown in Fig. 4 (a). At first, two pairs of SRRs with different side lengths are introduced on both sides of CWR without $VO_2$, and the coupling between the bright mode and dark modes with two different resonant frequencies will introduce two transparent peaks, the simulated and calculated transmission spectra are given in Fig. 4 (b).

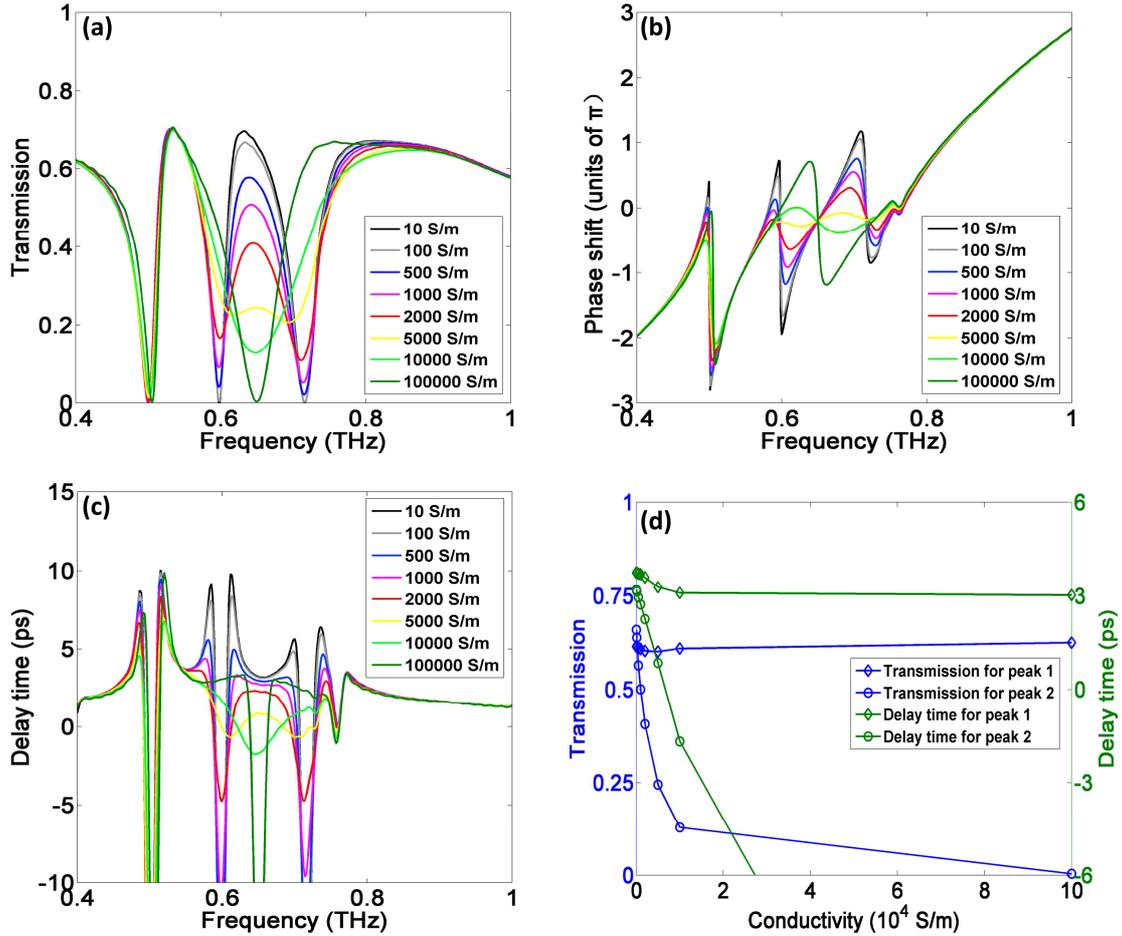

Fig. 5 (a) The transmission spectra of double-peak EIT under different conductivities of $VO_2$ embedded in SRRs on the left side of CWR. (b) The phase shift under different conductivities of VO2. (c) The delay time under different conductivities of VO2. (d) The transmission as well as delay time at 0.56 and 0.65 THz under different conductivities.

Then, $VO_2$ is embedded in SRRs on the left side of CWR. When the conductivity of $VO_2$ rises, peak 2 will drop, while the amplitude of peak 1 will remain unchanged, as shown in Fig. 5(a). At the same time, the variation of the phase shift as well as the group delay is also accompanied as depicted in Fig. 5(b, c). Clearly, the dispersion modulation is mainly introduced within the right transparent window. The transmission and delay time within the two transparent peaks, at 0.56 THz and 0.65 THz, are plotted under different conductivities as given in Fig. 5(d). For the peak 2, the transmission at 0.65 THz can be switched from 66% to 0.36% with the MD 99.45%, and the delay time can be tuned from 3.20 ps to be negative. However, for the peak 1, these values are almost stable. The transmission at 0.56 THz remain at 62%, and delay time changes from 3.76 ps to 3.03 ps merely.

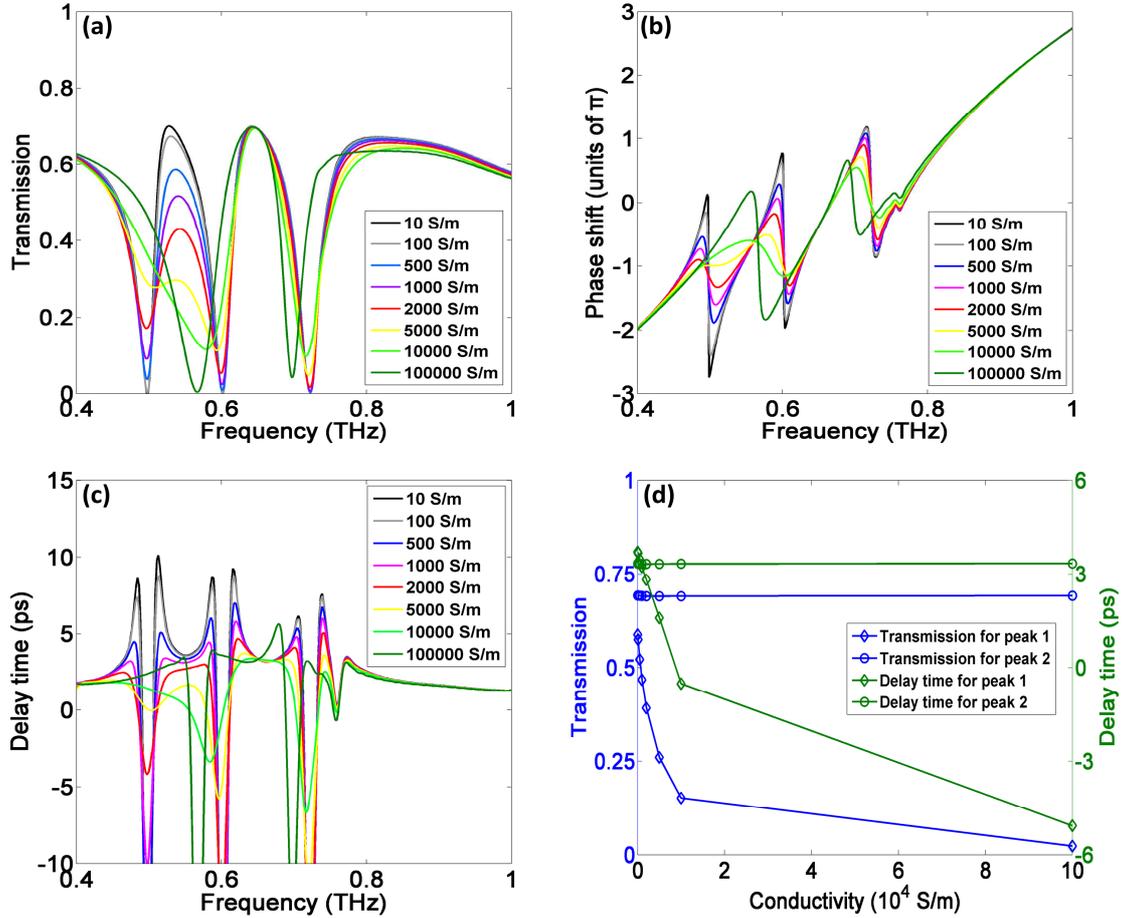

Fig. 6 (a) The transmission spectra of double-peak EIT under different conductivities of $VO_2$ embedded in SRRs on the right side of CWR. (b) The phase shift under different conductivities of VO2. (c) The delay time under different conductivities of $VO_2$. (d) The transmission as well as delay time at 0.56 and 0.65 THz under different conductivities.

Similarly, if VO2 is embedded in SRRs on the right side of CWR, the intensity of peak 1 as well as the dispersion within it will be tuned, while peak 2 will remain stable. The corresponding transmission spectra, phase shift and delay time with different conductivities of $VO_2$ are shown in Fig. 6 (a-c), the detailed values of transmission as well as delay time at 0.56 THz and 0.65 THz are plotted in Fig. (d). Expectedly, the transmission can be modulated from 59% to 2.3% with the MD 96.1% at 0.56 THz, and the delay time varies from 3.72 ps to be negative. Meanwhile, the transmission and delay time at 0.65 THz maintain at 69% and 3.34 ps, respectively. Therefore, the amplitudes of two peaks as well as the group delay within two transparent windows can be independently tunable utilizing $VO_2$. More importantly, if $VO_2$ islands are embedded into all four SRRs, the two peaks can be controlled both simultaneously and independently, as long as technically realizing separate stimulation of $VO_2$ islands on each side of CWR. What's more, EIT with more peaks can be realized by introducing more different coupled dark modes [47], and each peak can also be independently controlled using the strategy described above. Obviously, such structure can provide double switchable THz channels, and can play an important role in THz active devices.

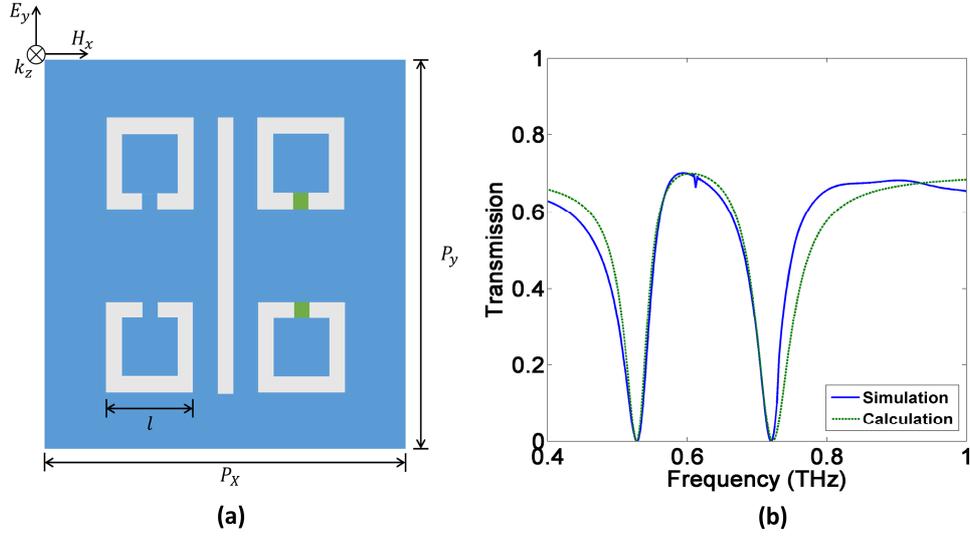

Fig. 7 (a) The structure to realize broadband EIT. The geometric parameters are unchanged except $l$= 29 um, $P_x$= 116 um and $P_y$= 120 um. (b) The simulated and calculated transmission spectra of broadband EIT.

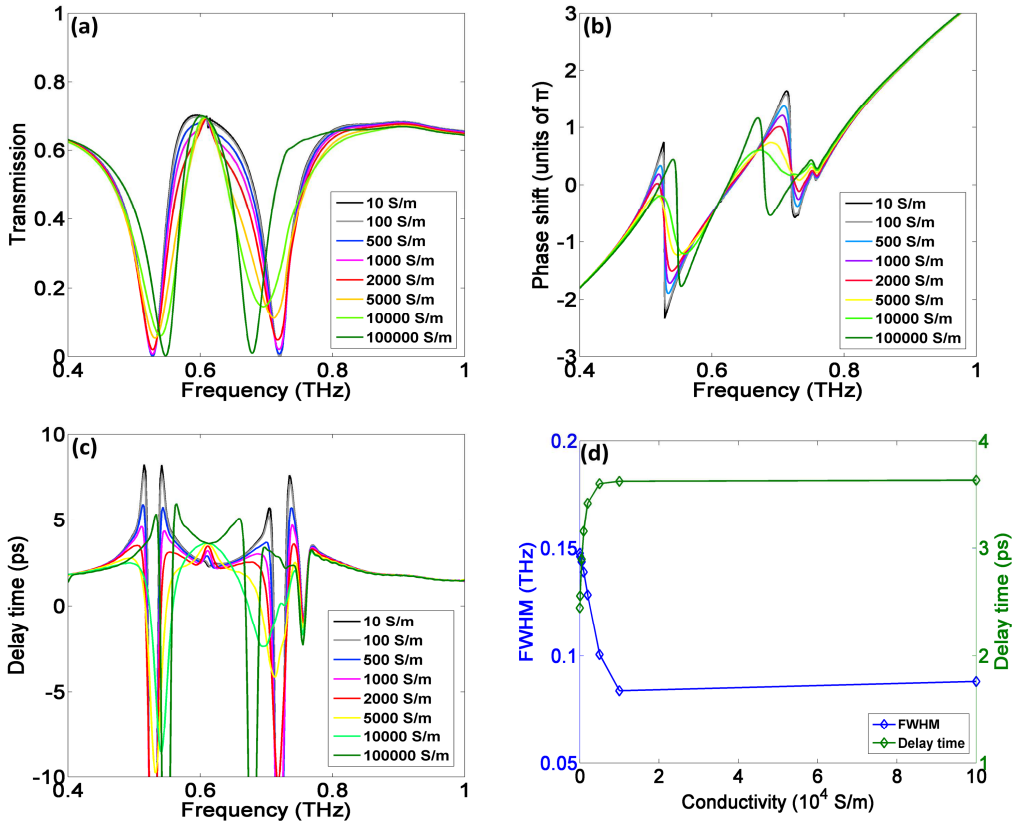

Fig. 8 (a) The transmission spectra of broadband EIT under different conductivities of $VO_2$ embedded in SRRs on the one side of CWR. (b) The phase shift under different conductivities of VO2. (c) The delay time under different conductivities of VO2. (d) The FWHM as well as delay time at 0.61 THz under different conductivities.

Finally, we investigate EIT with tunable quality (Q) factor based on the framework presented in Fig. 7(a). At first, the broadband EIT is realized by coupling four ideal SRRs without $VO_2$ beside the CWR, the simulated and calculated transmission spectra are shown in Fig. (b). Next, via embedding VO$_2$ into the gaps of SRRs on the one side of CWR, the Q factor of EIT resonance can be tuned as shown in Fig. 8(a), the corresponding phase shift and delay time dynamics are also provided in Fig. 8(b, c). The group delay at the center transparent frequency 0.61 THz and the width at half maximum (FWHM) of the transparent peak under different conductivities of VO$_2$ are shown in Fig. 8(d). With the increasing conductivity, the FWHM can be narrowed from 0.15 THz to 0.09 THz, thus the Q factor, defined as $Q = \omega/\text{FWHM}$, is switched from 4.07 to 6.78. Meanwhile, the transmission at center frequency is unchanged. Besides, the delay time increases from 2.45 ps to 3.64 ps at the cost of the reduction of effective slow light bandwidth. This is normal since a sharper peak of EIT indicates a more intense dispersion, which will bring about more significant slow light effect. This structure provides a way to modulate the bandwidth of transmitted waves as well as the effective slow light region. However, there is a trade-off between the bandwidth and intensity of the slow light effect.

## 4. Theoretical calculation

The MM-based EIT analogue can be quantitatively described by a coupling harmonic oscillator model [48,49]. We start from the double-peak EIT case, which can be described by a three-oscillator model. In this model, the CWR is seen as the oscillator 1 excited by the external electric field $E(t)$. The two pairs of SRRs are represented by two oscillators 2 and 3, which are driven only via the coupling with oscillator 1. The coupled differential equations are as follows:

$$\ddot{Q}_1(t) + \gamma_1 \dot{Q}_1(t) + \omega_1^2 Q_1(t) + \kappa_{12} Q_2(t) + \kappa_{13} Q_3(t) = gE(t)$$
$$\ddot{Q}_2(t) + \gamma_2 \dot{Q}_2(t) + \omega_2^2 Q_2(t) + \kappa_{12} Q_1(t) = 0 \quad (3)$$
$$\ddot{Q}_3(t) + \gamma_3 \dot{Q}_3(t) + \omega_3^2 Q_3(t) + \kappa_{13} Q_1(t) = 0$$

where $Q_1$, $Q_2$ and $Q_3$ are amplitudes of three oscillators, $\omega_1$, $\omega_2$ and $\omega_3$ are resonant frequencies of three oscillators, $\gamma_1$, $\gamma_2$ and $\gamma_3$ are damping factors of three oscillators, $\kappa_{12}$ is the coupling coefficient between oscillator 1 and 2, and $\kappa_{13}$ is the coupling coefficient between oscillator 1 and 3. Therefore, the energy dissipation of this system is obtained as follows:

$$P(\omega) \propto \frac{(\omega_2 - \omega - i\frac{\gamma_2}{2})(\omega_3 - \omega - i\frac{\gamma_3}{2})}{(\omega_2 - \omega - i\frac{\gamma_2}{2})(\omega_1 - \omega - i\frac{\gamma_1}{2})(\omega_3 - \omega - i\frac{\gamma_3}{2}) - \frac{\kappa_3^2}{4}(\omega_2 - \omega - i\frac{\gamma_2}{2}) - \frac{\kappa_2^2}{4}(\omega_3 - \omega - i\frac{\gamma_3}{2})} \quad (4)$$

The profile of EIT transmission spectrum can be fitted utilizing the equation (4), and all the calculated results in this paper are obtained by this method. In the double-peak EIT case, $\omega_2$ and $\omega_3$ are considered to be different. In the broadband EIT case, $\omega_2$ and $\omega_3$ are the same. In the single-peak EIT case, oscillator 3 is absent, thereby $Q_3$, $\omega_3$, $\gamma_3$ and $\kappa_{13}$ are all equal to zero.

## 5. Conclusion

In summary, EIT analogue based on a VO$_2$-assisted MM in THz band is proposed. Utilizing the insulator-to-metal transition of VO$_2$, we achieve diverse EIT modulations, including amplitudes, group delay, number of peaks and Q factor modulations of EIT. Under optical stimulation, sub-ps response time can be expected in these modulation processes. This work introduces active control of EIT with more degree of freedom, and can find potential applications in future wireless and ultrafast THz communication systems as multi-channel filters, switches, spacers, logic gates and modulators.

**Acknowledgments**

This work is supported by the National Natural Science Foundation of China (61671455, 61805278), the Foundation of NUDT (ZK17-03-01), the Program for New Century Excellent Talents in University (NCET-12-0142), and the China Postdoctoral Science Foundation (2018M633704).